\documentstyle[aps,preprint]{revtex}
\begin{document}
\preprint{PAR-LPTHE/95-17, EHU-FT/9506}
\draft
\tighten
\title{Multistring solutions in inflationary spacetimes}
\author{H.J. de Vega\cite{emLPTHE}}
\address{Laboratoire de Physique Th\'eorique et Hautes \'Energies
\\
Universit\'es Paris VI-VII - Laboratoire associ\'e au CNRS n$^{\rm o}$ 280
\\
Tour 16, 1er. \'et., 4, Place Jussieu, 75252 Paris Cedex 05
\\
FRANCE}
\author{I.L. Egusquiza\cite{emUPV}}
\address{Dpt. of Theoretical Physics\\
University of the Basque Country\\
Apdo. 644 P.K. - 48080 BILBAO\\
SPAIN}
\date{\today}
\maketitle
\begin{abstract}
Multistring solutions of the string equations of motion
are found
for inflationary spacetimes with expansion factor $R(\eta)\propto
\eta^k$ for any  $k<0$  and $\eta$ conformal
time. If $0>k>-1$ only two-string solutions can be found within our ansatz,
whereas for $k=-1-1/n$, with $n=1,2,\ldots$, multistring solutions exist with
an
infinite number of strings [In the special case
$k=-2$  we recover de Sitter spacetime where multistring solutions were first
found].
\end{abstract}
\section{Introduction and Motivations}
The study of string dynamics in curved spacetime
reveals new insights and
new physical phenomena with respect to string propagation in flat spacetime
\cite{plb}-\cite{erice}.
The results of
this programme are relevant both for fundamental (quantum) strings and for
cosmic strings, which behave essentially in a classical way.

Recently, a novel feature for strings in de Sitter spacetime
was found: exact multi-string solutions
\cite{dvmisa}-\cite{hnamulti}
The novel feature
 is that one single world-sheet gives multiple (infinitely many)
strings. The world-sheet time $\tau$ turns out to be a multiply-valued
\cite{dvmisa}
(and even an infinite-valued \cite{hnamulti})
function of the target string time $X^0$ (which can be
 the comoving or cosmic time $t$ or the conformal time $\eta$).
Each branch of $\tau$ as a
function of $X^0$ corresponds to a different string.
In flat spacetime,
multiple string solutions are necessarily described by multiple
world-sheets. Here, a single world-sheet describes many different
and simultaneous strings as a consequence of the coupling with the
spacetime geometry. These strings do not interact among themselves; all the
interaction is with the curved spacetime.

In the present note we show that multistring solutions are present in universes
with  scale factor $R(\eta)\propto \eta^k$ for $k$ a real number. [Here $\eta$
stands for conformal time]. Within our ansatz the multistring solutions for
$0>k>-1$ are shown to contain at most two strings, as opposed to the case
$k<-1$,
for which the possibility of having an infinite number of strings exists.

\section{Equations of motion.}
Let us consider loop strings moving in cosmological background with metric
$${\rm d}s^2=R^2(\eta)\left({\rm d}\eta^2-{\rm d}{\vec X}^2\right)\,,$$
where $\eta$ is the conformal time, ${\vec X}$ stands for the vector of spatial
coordinates, and $R(\eta)$ is the conformal factor, which we assume takes
the form $R^2(\eta)=\eta^k$. For $k > 0$ we have FRW spacetimes:
$k=4$ is the matter-dominated universe, $k=2$ is the
radiation-dominated  universe. For  $k < 0$ we have inflationary spacetimes:
$-2<k<0$ corresponds to  superinflation,  $k=-2$ is de Sitter
spacetime and $k<-2$  corresponds to power inflation.

The following Ansatz for the string coordinates
\begin{eqnarray}
\eta & = & \eta(\tau)\,,\nonumber\\
X^1 & = & f(\tau)\cos\sigma\,,\nonumber\\
X^2 & = & f(\tau)\sin\sigma\,,\nonumber\\
X^3 & = & 0\, .  \label{ansatz}
\end{eqnarray}
separates the string  equations of motion and  string constraints
\cite{hjile,letelier}. Here,   $\tau$ and  $\sigma$ stand for the worldsheet
coordinates. That is, a set of ordinary differential equations follows
for $ \eta(\tau) $ and $ f(\tau)$:
\begin{eqnarray}
\eta\ddot\eta + k\dot f^2 &=0\,,\nonumber\\
\eta\ddot f + k \dot\eta\dot f + \eta f &=0\,,\nonumber\\
\dot\eta^2 - \dot f^2 - f^2 & = 0\,. \label{ecuaciones}
\end{eqnarray}
The string proper size  under this Ansatz becomes $S(\tau)=\eta^{k/2} f$.
The  string energy and pressure are given by
\begin{equation}
E(\tau)=\eta^{k/2}\dot\eta\quad , \quad P={{\eta^{k/2}}\over{2\,
\dot\eta}}(\dot f^2-  f^2) \; ,\label{energia}
\end{equation}
when we set the string tension to 1.

Examining equations (\ref{ecuaciones}), we observe that the third one
(the constraint) is a
conserved  quantity of the first two. Furthermore, the system is autonomous. To
top it all, the system is homogeneous in $\eta,f$. This means that we can
reduce eqs. (\ref{ecuaciones}) to one single nonlinear differential equation,
plus two quadratures \cite{journees,letelier}.

Let us make the change of variables
\begin{equation}
z:={f\over\eta}\,,\qquad v(z):={{\dot f}\over{\dot\eta}}\,,
\label{change}
\end{equation}
which leads to the equation
\begin{equation}
z(v-z){{{\rm d}v}\over{{\rm d}z}} + (1+ kzv)(1-v^2) =0 \,.\label{ecuared}
\end{equation}
Note that for the de Sitter universe ($k=-2$), $z$ is the proper size
of the string,
and $v$ is given in terms of proper size and energy as
$v=S+\dot S/E$.

For general $k$, we have
\begin{eqnarray}
v^2 & = & 1- {{S^2}\over{E^2}}\,,\nonumber\\
v &=& -{k\over2}z + {{\dot S}\over E} \,.\nonumber
\end{eqnarray}

The  change of variables can be inverted as
\begin{eqnarray}\label{etaz}
\eta(z)&=&\eta(z_0)\exp\left(\int_{{z_0}}^z{{{\rm
d}z}\over{v(z)-z}}\right)\,, \\
f(z) &=& z \eta(z) \cr
\dot z^2&=&{{(v-z)^2 z^2}\over{1-v^2}}\,.
\label{zetatau}
\end{eqnarray}

{F}rom this last equation we immediately see that in order to have a physical
solution $v^2$ must be smaller than one. This requirement can be avoided if we
allow for imaginary $\tau$. The significance of such a relaxation is that we
would be admitting instantonic (or superluminal) strings. That is to
say, string solutions such that the worldsheet metric has signature
$(-,-)$, as opposed to the lorentzian one $(+,-)$.

A symmetry of equations (\ref{ecuared}) and (\ref{zetatau}) is given by the
simultaneous change $z\to-z$ and $v\to-v$. We can therefore concentrate on
either $z>0$ or $z<0$. We shall take $z>0$. The physical strip is then
\begin{equation}
\{(z,v)\mid z\geq0 \quad{\rm and}\quad-1<v<1\}\,.
\end{equation}

{F}rom equation (\ref{ecuared}) we see that there will be singular
points in the
$(z,v)$ plane given by the intersection of the lines
$$v=-{1\over{kz}}\,,\quad {\rm and}\quad v=\pm1\,,$$
with the lines
$$v=z\,,\quad {\rm and}\quad z=0\,.$$
That is to say, the singular points will be
$$(0,\pm1)\,,\quad(\pm1,\pm1)\,,
\quad(\pm{1\over{\sqrt{-k}}},\pm{1\over{\sqrt{-k}}})\,,$$ The last two
singular points are real only  for $k<0$, and only for $k<-1$ is one
of them to be
found in the physical strip. They correspond to exact solutions of eqs.
(\ref{ecuaciones}) which were called  instantons in Ref. \onlinecite{hjile}.

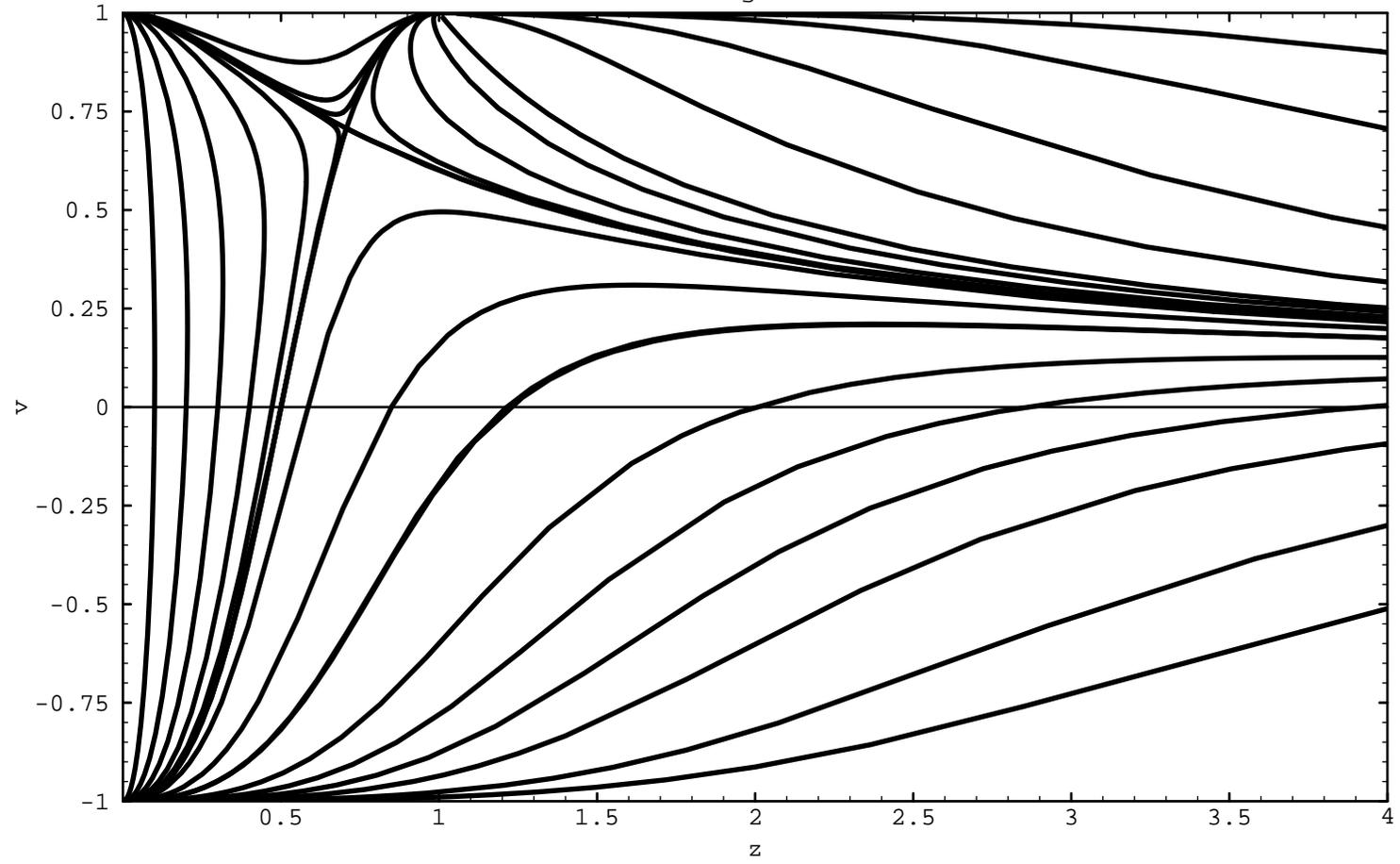
\begin{figure}
\caption{Physical strip of
the phase portrait of equation (\protect\ref{ecuared}) for de Sitter
spacetime.}
\label{fig1}
\end{figure}

In fig. \ref{fig1} we portray the physical strip for de Sitter
spacetime ($k= -2$). The
saddle point at $({{1}\over{\sqrt{2}}},{{1}\over{\sqrt{2}}})$ is the singular
point corresponding to the instanton, which was called $q^o$ in Ref.
\onlinecite{dvmisa}.
The physical strip for the case $k=-3/2$ is portrayed in fig. \ref{fig2}.
In the next figure, fig. \ref{fig3}, only the separatrices for de Sitter are
shown.

Each singular point corresponds to asymptotic or exact solutions of eq.
(\ref{ecuaciones}), as was displayed in Ref. \onlinecite{journees}.
We shall now
analyze these asymptotics in order to know whether the   singular points
are reached in finite
string time ($\tau$). Recall that the presence of multistring
solutions arises from the multivaluedness of $\tau(\eta)$.
Since, if $\tau(\eta)$ is multivalued, it follows that the
physical range of $\eta$, that is, $(0,\infty)$ must correspond to a finite
range in $\tau$. Since it is clear that only the aforementioned singular points
can correspond to the singular values  of $\eta$, it follows that a necessary
precondition for the presence of multistring configurations is that some of the
graphs of the solutions to eq. (\ref{ecuared}) be given by a finite string
time.

Note that the string time $\tau$ is obtained
from the solutions of eqs. (\ref{ecuared}) through (\ref{zetatau}) as
\begin{equation}
\tau-\tau_0=\int^{z(\tau)}_{{z_0}}\,{\rm
d}z\,{{(1-v^2)^{1/2}}\over{(v-z)z}}\,.\label{integraltau}
\end{equation}

\section{Behaviour near the singular points}

Let us briefly summarize the behaviour of $\eta(\tau), f(\tau)$ and
$v(z)$ near the singular points $(z,v)=(0,\pm1)\,,\quad(\pm1,\pm1)\,,
\quad(\pm{1\over{\sqrt{-k}}},\pm{1\over{\sqrt{-k}}})$
\cite{hjile} -\cite{journees} -\cite{backr}.

The points $(z,v)=(0,\pm1)$ correspond to the string behaviour
\begin{eqnarray}\label{eta0}
\eta(\tau) &\buildrel {\tau\to\tau_0} \over =&\eta_0 + {{\eta_0}\over
{a\sqrt{2}}}\,(\tau-\tau_0) + O(\tau-\tau_0)^3 ,\cr
f(\tau) &\buildrel {\tau\to\tau_0} \over =&\pm{{\eta_0}\over
{a\sqrt{2}}}\,(\tau-\tau_0) + O(\tau-\tau_0)^3 ,\cr
v  &\buildrel {z \to 0}  \over =& \pm (1 - a z^2) ,\cr
P &\buildrel {\tau\to\tau_0} \over=& {1\over 2}\,E\buildrel
{\tau\to\tau_0} \over = {{\eta_0^{k/2+1}}\over {2a\sqrt{2}}}.
\end{eqnarray}
where $a>0$.

The relation between $P$ and $E$ is like in the case of radiation.
It is similar to the dual to unstable behaviour although here  $P$ and $E$ tend
to {\em finite} constants. Eqs.(\ref{eta0}) hold for all values of $k$.

Near the point  $(z,v)=(1,1)$ we find
\begin{eqnarray}\label{unouno}
\eta(\tau) &\buildrel {\tau\to\tau_1} \over =&
(\tau-\tau_1)^{1/(k+1)} \left[ 1 + {k \over 6} \,
 (\tau-\tau_1)^2 + O (\tau-\tau_1)^3 \right] ,\cr
f(\tau) &\buildrel{\tau\to\tau_1} \over =&(\tau-\tau_1)^{1/(k+1)}
\left[ 1 - {{k^2 + 2 k + 3} \over {6(2k+3)}}\,  (\tau-\tau_1)^2 \right. \cr
& & \left.  + \;  O(\tau-\tau_1)^3  \; \right] ,\cr
v  &\buildrel {z \to 1}  \over =& 1 + (2k+3)(z-1) ,\mbox{~for~}k>-3/2 \cr
v  &\buildrel {z \to 1}  \over =& 1 - A (1-z)^{-2k-2} , \;\;
\mbox{for~}k<-3/2 \cr
P &\buildrel {\tau\to\tau_1} \over=& {1\over 2}\,E\buildrel
{\tau\to\tau_1} \over = {1 \over {2(k+1)}}
(\tau-\tau_1)^{-{k\over {2(k+1)}}}  .
\end{eqnarray}
where $A>0$.

We find here dual to unstable behaviour for $k>0$ and for $k<-1$. For
$-1<k<0$, the equation of state is the same (radiation) but  $P$ and $E$ tend
to zero.

Near $(z,v)=({1\over{\sqrt{-k}}},{1\over{\sqrt{-k}}})$, we find for $k<-1$,
\begin{eqnarray}\label{menok}
\eta(\tau) &\buildrel {\tau\to+\infty} \over =& e^{\mp\sqrt{-k-1}\tau}\cr
f(\tau) &\buildrel {\eta^{\pm1}\to 0} \over =&{{\eta}\over \sqrt{-k}}
+ a \, \eta^{\lambda_{\pm}}\cr
v &\buildrel {z \to {1\over \sqrt{-k}} }  \over =&  {1\over \sqrt{-k}} +
\lambda_{\pm} \left(z - {1\over \sqrt{-k}}\right)\cr
P &\buildrel {\tau\to\tau_1} \over=& {1\over 2}\, E\buildrel
{\tau\to\tau_1} \over = {1 \over {2 \sqrt{-k-1}}}\, \eta^{k/2+1}
\end{eqnarray}
Here, $\lambda_{\pm}\equiv -{k \over 2} \pm {{\sqrt{k^2-4k-4}}\over 2}
$. We find here dual to unstable behaviour for $k<-2, \eta \to 0$ and
for $-2<k<-1, \eta \to \infty$. Otherwise,  the equation of state is
the same (radiation) but  $P$ and $E$ tend to zero.

Let us now consider the asymptotic points  $(z,v)=(+\infty,\pm 1)$ and
$(z,v)=(\infty,0)$.

We find for  $(+\infty,\pm 1)$ and $k>0$,
\begin{eqnarray}\label{infuno}
\eta(\tau) &\buildrel {\tau\to\tau_2} \over =& (\tau-\tau_2)^{1/(k+1)},\cr
f(\tau) &\buildrel {\tau\to\tau_2} \over =& f_0 \pm
(\tau-\tau_2)^{1/(k+1)} ,\cr
v &\buildrel {z \to +\infty}  \over =& \pm\left[ 1 - {a \over {z^k}}
\right] \cr
P&\buildrel {\tau\to\tau_2} \over =& {1\over 2}\, E\buildrel
{\tau\to\tau_2} \over = {1 \over {2(k+1)}}\,  (\tau-\tau_2)^{-{k\over
{2(k+1)}}}\to \infty .
\end{eqnarray}
where $a>0$. Dual to unstable behaviour appears again.

For  $(+\infty, 0)$, where the separatrices  end for $k>0$ and $k<-1$, we have
\begin{eqnarray}\label{infcero}
\eta(\tau) &\buildrel{\tau\to\tau_3}\over=& \tau-\tau_3 ,\cr
f(\tau) &\buildrel
{\tau\to\tau_3} \over =& 1 - {{(\tau-\tau_3)^2}\over{2(k+1)}} ,\cr
v  &\buildrel {z \to +\infty}  \over =& -{1 \over {(k+1)z}}\to 0 ,\cr
P  &\buildrel
{\tau\to\tau_3} \over =& -{1\over 2}E \buildrel
{\tau\to\tau_3} \over  = -{1 \over 2}  (\tau-\tau_3)^{k/2}.
\end{eqnarray}
We find here the equation of state of unstable strings. However,  $P$
and $E$ tend to zero for $k>0$ instead of blowing up.

When $k$ is in the interval $-1<k<0$ a different behaviour appears
near  $(+\infty, 0)$,
\begin{eqnarray}\label{incekn}
\eta(\tau) &\buildrel{\tau\to\tau_3}\over=& \tau-\tau_3 ,\cr
f(\tau) &\buildrel
{\tau\to\tau_3} \over =& 1 + b  (\tau-\tau_3)^{1-k}\cr
v  &\buildrel {z \to +\infty}  \over =& \pm a z^k \cr
P  &\buildrel
{\tau\to\tau_3} \over =& -{1\over 2}E \buildrel
{\tau\to\tau_3} \over  = -{1 \over 2}  (\tau-\tau_3)^{k/2}\to -\infty
\end{eqnarray}
Here we find unstable behaviour.

In  Eqs.(\ref{eta0})-(\ref{incekn}), $\tau_0,\ldots, \tau_3 $ are arbitrary
real constants.

For any value of $k$ and large $\tau$, the ring strings may exhibit the
oscillatory behaviour\cite{hjile}
\begin{eqnarray}
\eta(\tau)&\buildrel{{\tau\to+\infty}}\over=
&\tau^{2/(k+2)}\,,\nonumber \\
f(\tau)&\buildrel{{\tau\to+\infty}}\over= &{2\over{k+2}} \; \tau^{-k/(k+2)}
\cos(\tau+\varphi)\,,\label{solasim}\\
\nonumber
\end{eqnarray}
where $ \varphi $ is a constant phase and the oscillation amplitude has been
 normalized. For large $\tau $, the energy and pressure of the
solution (\ref{solasim}) show stable behaviour
\begin{eqnarray}
E(\tau) & \buildrel{{\tau\to+\infty}}\over= & {{2 } \over {
k+2}} =  {\rm constant~} \quad, \nonumber \\
P (\tau) & \buildrel{{\tau\to+\infty}}\over= & -{1 \over {
k+2}} \cos(2\tau + 2\varphi) \to 0 \quad .
\end{eqnarray}
In the $(z,v)$ variables this corresponds to
\begin{eqnarray}
z&\buildrel{{\tau\to+\infty}}\over=&{2\over{k+2}}\,
{{\cos(\tau+\varphi)}\over{\tau}}\cr
v &\buildrel{{\tau\to+\infty}}\over=&- \sin(\tau+\varphi) \quad .
\end{eqnarray}
This represents cycles wound around the segment $-1<v<+1$ with a
width in the $z$ direction of order

$$1/[\tau(1+k/2)].$$

\section{Phase portrait}
We shall divide this section according to the bifurcation structure of eq.
(\ref{ecuared}), i.e., $k>0$, $0>k>-1$, and  $-1>k$.
\subsection{$k>0$}
For $k>0$ there are three kinds of solutions in the physical strip, to which
two separatrices are to be added. Namely, solutions joining
\begin{itemize}
\item[i)] $(+\infty,-1)$ with $(0,-1)$;
\item[ii)] $(+\infty,+1)$ with $(0,-1)$;
\item[iii)] $(0,-1)$ with $(0,1)$,
\end{itemize}
and the separatrices from $(0,-1)$ to $(1,1)$ and from $(0,-1)$ to
$(+\infty,0)$. From the asymptotic behaviour of $v(z)$ at the singular points,
we see that solutions of type (i) and (ii) require a finite interval of $\tau$.
That is, a finite $\tau=\tau_i$  corresponds to the initial situation
$(+\infty,\pm1)$, and after a finite string time $\tau_f$ we are to find
ourselves at the point $(0,-1)$.

These solutions correspond to strings that originate at the
big bang [$R=0, \eta=0, (z,v)=(+\infty,\pm 1)$],
that collapse to zero size at $(0,-1)$ in a
finite cosmological time.  The posterior oscillatory
behaviour of the worldsheet is given by type (iii) solutions, for which the
$\tau$ interval is also finite. The initial, type (i) or (ii) behaviour give
rise to a unique type (iii) solution:
equation  (\ref{ecuared}) is singular at the point $(0,-1)$. Therefore, the
solution is not unique: it is not determined by the initial value $v(0)$. As a
matter of fact, it follows from the equation that for consistency $v'(0)=0$. So
the solutions to the equation are characterized by the second derivative of $v$
with respect to $z$ at the point $z=0$: $v\sim-1+a  z^2$ [see
eq.(\ref{eta0})].  For the physical
reasons previously mentioned, it is clear that $a>0$. After some
manipulations, we can relate $a$ to physical quantities, namely the
scale factor at this point, $R_0$, and the energy of the string, $E_0$, at
this same instant, by use of the expression
$$a={{R_0^{1+2/k}}\over{2E_0^2}}\,.$$
It follows that the solutions which reach this point are uniquely
characterized
by this physical quantity (here, by solutions we mean solutions in the $(z,v)$
plane, which can each be mapped to a class of equivalent physical solutions for
$\eta$ and $f$). Moreover, by the simultaneous mapping $v\to-v$ and
$z\to-z$,
we
can identify the solution in the physical strip that corresponds to an ingoing
one with the calculated $a$, and we see that for the solutions coming from
$(+\infty,-1)$ to the point $(0,-1)$ the continuing solution is of the
oscillatory form iii).

Therefore, no multistring behaviour with infinite number of strings is found.
The reason being that all the solutions previously stated  require a
semiinfinite interval of string time, thus making it impossible to have more
than two strings present in each solution. Furthermore, in order to have even
these two strings, it would be necessary to tie them together at $\eta=0$. But
the behaviour (\ref{infuno}) forbids this; from which it follows that within
our ansatz no multistring solutions exist for $k>0$.

Let us now analyze the exceptional solutions, the separatrices.
Both of them
traverse a finite $\tau$ interval, and both of them correspond to strings
originating at the big bang which collapse later on. The difference with the
previously mentioned type (i) and (ii) solutions is that whereas
solutions (i) and (ii) exhibit dual to unstable behaviour
[see eq.(\ref{infuno})]
and  $S\sim R \to 0$ near $(+\infty,\pm 1)$,
for the $(1,1)$ to $(0,-1)$ separatrix
there is also dual to unstable behaviour [see eq.(\ref{unouno})] near
$(1,1)$ but there $ S\sim R^{1+2/k} \to 0$.
For the other one, $(+\infty,0)$ to $(0,-1)$,
there is unstable behaviour near  $(+\infty,0)$ and
$S \sim R \sim 1/E$ as $R\to0$.

\subsection{$0>k>-1$}
There are two kinds of solutions in the physical strip: $(0,-1)$ joined with
$(+\infty,0)$, and $(0,-1)$ joined to $(0,1)$. Additional to these, one finds a
separatrix $(0,-1)$ to $(1,1)$. The $\tau$ interval for the $(0,-1)$ to
$(+\infty,0)$ solutions is finite, and the strings thus described correspond
to [see eq.(\ref{incekn})]
 $R\sim S\sim E\sim z^{-k/2}\to\infty$ as $z\to+\infty$, with
$z \sim1/(\tau - \tau_3)$. For these
superinflationary spacetimes, there are then strings that start at
$t=0, [R=\infty] $ and
proceed to collapse, then continuing with an oscillatory motion (the second
class of solutions), similarly to what happens in the previous $k>0$ case.
The separatrix is also analogous to the previous case, and no multistring
solutions with an infinite number of strings are present in the ring solutions
to
the string equations of motion for these spacetimes. Nonetheless, the argument
against the existence of two-string solutions valid for $k>0$ fails in the case
at hand, and two-string solutions exist within our ansatz for $0>k>-1$. This is
depicted in figs. \ref{figdedos}. This is due to the fact that the behaviour
for $\eta\to0$ is now given by  eqns. (\ref{incekn}).

\subsection{$k<-1$}
There are four kinds of solutions to equation (\ref{ecuared}) in the physical
strip, and four corresponding separatrices. The solutions join
\begin{itemize}
\item[i)] $(+\infty,0)$ to  $(1,1)$,
\item[ii)] $(1,1)$ to $(0,1)$,
\item[iii)] $(0,-1)$ to $(0,1)$,
\item[iv)] $(0,-1)$ to $(+\infty,0)$,
\end{itemize}
with separatrices from the point $(1/\sqrt{-k},1/\sqrt{-k})$ to each of the
points $(0,\pm1)$, $(1,1)$, and $(+\infty,0)$.

The $(+\infty,0)$ point corresponds to $R\sim S\sim E\sim
z^{-k/2}\to\infty$, [see eq.(\ref{infcero})]
and is reached in finite $\tau$ time from the non-singular points
connected to it . The points $(0,\pm1)$, as before, correspond  to
finite $R$ and $E$,
and $S=0$. The points $(1,1)$ and $(1/\sqrt{-k},1/\sqrt{-k})$, however, are
more complicated.

As was mentioned before, the point $(1/\sqrt{-k},1/\sqrt{-k})$ is a solution
of the equations of motion of the string that we had previously called the
instanton for $k+1>0$\cite{hjile}.
For this solution, invariant string size and string energy are
both proportional to cosmological time.

The point $(1,1)$ is a node for $k<-1$. Writing eq. (\ref{ecuared}) as a
system of two differential equations, we obtain the linearized
solution close to
this point:
\begin{equation}\label{unun}
z\sim  1 - {{1-v}\over{2k+3}}+ \alpha \; (1-v)^{-1/(2k+2)}\,,
\end{equation}
with $\alpha$ some constant. It should be noticed that for $k=-3/2$ this
linearized solution involves a logarithmic term.
{F}rom this equation, or {from} a direct analysis of
eq. (\ref{ecuared}) we see that
the leading behaviour of $\eta(\tau)$  is given by $(\tau-\tau_0)^{1/(k+1)}$.
The value  $\infty$ is reached by $\eta$ in a finite time from the non-singular
points connected to $(1,1)$.

It is then clear that the interval $(0,\infty)$ of $\eta$ is obtained from a
finite interval of string time, this being the prerequisite for the presence
of multistring configurations. The point $\eta=0$ corresponds to the
 singular point  $(1/\sqrt{-k},1/\sqrt{-k})$ and to
$(z,v)=(+\infty,0)$.  Disregarding the
separatrix for the moment, consider a string that comes out of $\eta=0$ with
positive ${\rm d}f/{\rm d}\eta$ (i.e., from $(z,v)=(+\infty,0^+)$). By
following the evolution in the physical strip, we see it eventually reaches
$(z,v)=(1,1)$, in finite string time.

Similarly if the starting point is $(z,v)=(+\infty,0^-)$: In its evolution
in the $(z,v)$ plane, the string moves towards the point $(0,-1)$. By the usual
identification, this is the same as the point $(0,1)$, from which it comes out,
and then either goes on to an oscillatory behaviour, as before, or is driven to
the point $(1,1)$, which is again reached in a finite time.

In what concerns the  point $(1/\sqrt{-k},1/\sqrt{-k})$, it can only
be reached via the
separatrices. It should be observed that it is a saddle point, and there are
then two separatrices corresponding to $\eta\to0$ as we approach this singular
point, and another two such that $\eta\to\infty$. So, if we were to start from
$(z,v)=(\infty,0)$, we would reach the  point
$(1/\sqrt{-k},1/\sqrt{-k})$ through the more
``horizontal" separatrices (either going through the points $(0,\pm1)$ or not).
This would happen in infinite string time.
On the other hand, the separatrix connecting the  singular point
$(1/\sqrt{-k},1/\sqrt{-k})$
with $(1,1)$ corresponds to $\eta=0$ as we approach
$(1/\sqrt{-k},1/\sqrt{-k})$, and
$\eta\to\infty$ as we approach $(1,1)$. This also needs infinite
string time $\tau$ to take place.

Notice that an infinite string time $\tau$ is needed to reach the point
$(1/\sqrt{-k},1/\sqrt{-k})$. This is the only point which such property.

\section{Multistring solutions} The
requirement for a multistring configuration is that both $\eta(\tau)$ and
$X^0(\tau)$ be  functions of $\tau$ such that
\begin{enumerate} \item the only singularities correspond to the spacetime
singularity; \item the infinite range of $\eta$ ($X^0$) is obtained from a
finite range of string time. \end{enumerate}

These statements are well behaved under both worldsheet and spacetime
diffeomorphisms.

In principle, the very fact that a finite $\tau$ interval is mapped into the
whole physical range allows us to write multistring configurations, and we have
already shown some two-string solutions. For instance,
in fig. \ref{fig4} we portray a sequence of
graphics for the case
$k=-1.6$. First comes $|\eta|$ as a function of $\tau$. Note in this respect
that only the absolute value of $\eta$  is relevant here; and,
furthermore, that eqs. (\ref{ecuaciones}) are invariant under the mapping
$\eta\to-\eta$.  It should also be remarked that since the interval in $\tau$
is
finite, more strings can be accommodated onto the same worldsheet. Fig.
\ref{fig4}b shows
$f$ as a function of
$\eta$ for this solution: it is clear that at least two strings are present at
the same time. Both strings start at $\eta=0$. Their radius is the same only
at $\eta=0$, as depicted. To give a more complete picture, we next display
$v(z)$ for this solution.

The question remains posed, whether
something similar to what has been achieved in de Sitter can be transposed to
the more general context. In order to understand the problem better we shall
study the results of
\cite{dvmisa},\cite{hnamulti}, translated into our
formulation. As we have already pointed out, in de Sitter spacetime $z$ is the
proper size of the string,
and $v=\sqrt{1-S^2/E^2}$, so we can use the expressions given in
\cite{dvmisa} to obtain solutions to our equation (\ref{ecuared}). In
particular, the  hyperbolic solutions $q_-$ and $q_+$
(\cite{dvmisa}) are given by
\begin{equation}
v_{II}(z)={{2z-1}\over{2z^2 - 2 z +1}}\,.
\end{equation}
When we make use of the invariance under $z\to-z$, $v\to-v$, we see that this
solution corresponds exactly to the separatrices in the physical strip, as
shown in fig. \ref{fig3}, by considering also the solution
\begin{equation}
v_I(z)={{2z+1}\over{2z^2 + 2 z +1}}\,.
\end{equation}
Notice that $z = {1 \over {\sqrt{2}}}\,\coth{{\tau}\over{\sqrt{2}}} $
for the solution  $q_-$ and  $z = {1 \over
{\sqrt{2}}}\,\tanh{{\tau}\over{\sqrt{2}}} $ for the solution
$q_+$. Therefore, $z \geq {1 \over {\sqrt{2}}}$ corresponds to  $q_-$
and  $z \leq {1 \over {\sqrt{2}}}$ to $q_+$.

These two solutions, taken together, describe a total of three
strings.
One
string solution corresponds to starting at the  point
$(1/\sqrt{2},1/\sqrt{2})$ and following
$v_{II}$ downwards and to the left until it reaches $(0,-1)$, whence,
by the usual mapping, we
jump to $(0,1)$ and follow $v_I$ until the  point
$(1/\sqrt{2},1/\sqrt{2})$ is
reached from the left. This requires infinite string time. Another
string solution is given by the rest of $v_{II}$. That is
starting at the
 point $(1/\sqrt{2},1/\sqrt{2})$ upwards until the point $(1,1)$ is reached (in
infinite
string time), and then continuing by
  the line joining  $(1,1)$ to $(+\infty,0)$, which
needs finite time to be traversed. A further string corresponds to the
separatrix from $(+\infty,0)$ to the  point $(1/\sqrt{2},1/\sqrt{2})$,
requiring infinite time, and which is given by $v_I$.

The primitive that corresponds to $(1-v_{I,II}^2)^{(1/2)}/z(v_{I,II}-z)$ is
$\sqrt{2}\,{\rm arctanh}(\sqrt{2}\,z)$, whence the previous statements follow.
Notice that for $z>1/\sqrt{2}$ the free integration constant must be a complex
number.

In \cite{hnamulti} another class of multistring solutions is analyzed. Their
explicit expressions in terms of elliptic functions are not illuminating for
our present purposes, but their asymptotic behaviour close to the point $(1,1)$
in the $(z,v)$ plane can be extracted, thus becoming amenable to
comparison with
eq. (\ref{unun}). The result is that $v(z)\sim 1- \beta^2 (1-z)^2+\ldots$ close
to $z=1$. On comparison with eq. (\ref{unun}) we see it tallies perfectly, as
it should. We then conclude that for this type of multistring solutions to be
present, we need some way of continuing solutions in the $(z,v)$ plane through
the singular point $(1,1)$.

Notice that this demand is not fulfilled by the already depicted two-string
solutions, and that this way of obtaining/identifying multistring solutions is
different from the continuation through $z=\infty$ (i.e. $\eta=0$) previously
considered.

We can obtain multistring solutions by continuing solutions in the $(z,v)$
plane
through the singular point $(1,1)$; that means computing a solution of eq.
(\ref{ecuared}) which is well behaved through the point $(1,1)$, and then use
eqs. (\ref{etaz}) and (\ref{integraltau}) with a judicious choice of
integration
constants for the two branches of the $(z,v)$ solution which come out of the
point $(1,1)$. In order to do this consistently it is required that there be
continuous, well-behaved solutions in the $(z,v)$ plane through the singular
point. From the leading behaviour in eqn. (\ref{unun})  we see that a necessary
condition for this construction to work is that $k=-1-n$: for these
spacetimes multistring solutions exist with more than two strings.

Another alternative for the consistent construction of multistring solutions is
given from the examination of eq.(\ref{unouno}), whence we see  that $\eta$ and
$f$ are real near
$(1,1)$ in both sides of the point  $\tau = \tau_1$ provided $1/(k+1)$ is an
integer. Hence we
find real multistring solutions for $k = - 1 -1/n$ where $n$ is a
natural number. For $n=1$ we recover de Sitter spacetime.

In addition, we find from eq.(\ref{unouno})  near  the point  $\tau =
\tau_1$,
\begin{equation}
v \buildrel {\tau\to\tau_1} \over = 1 - {1 \over 2} (k+1)^2\, (\tau -
\tau_1)^2 + O (\tau -\tau_1)^3
\end{equation}
This shows that the string always stays within the physical strip
$-1\leq v \leq +1$ when $\tau$ goes beyond $\tau_1$.
In summary, an extra string appears for each point  $\tau_1$
exhibiting the behaviour (\ref{unouno}) near $(1,1)$.

The conclusion holds for generic solutions of the equations of motion, so the
statement is even stronger than the mere existence of multistring solutions:
for spacetimes with $k=-1-1/n$ there exists a whole continuous class of
multistring solutions to the (classical) string equations of motion.

The conclusion is that multistring
solutions will always be allowed for spacetimes with scalar factor of the form
$R(\eta)\propto\eta^k$ with $k<0$. For $k<0$  two-string solutions will be
possible. For $k=-1-1/n$ an infinite number of strings can be present in
multistring solutions. For $k=-1-n$ multistring solutions are allowed with more
than two strings. Note that de Sitter spacetime is included in both series of
special spacetimes.
\acknowledgements
Thanks are due to the Department of Theoretical Physics of the University
of Zaragoza, where part of this work was carried out.


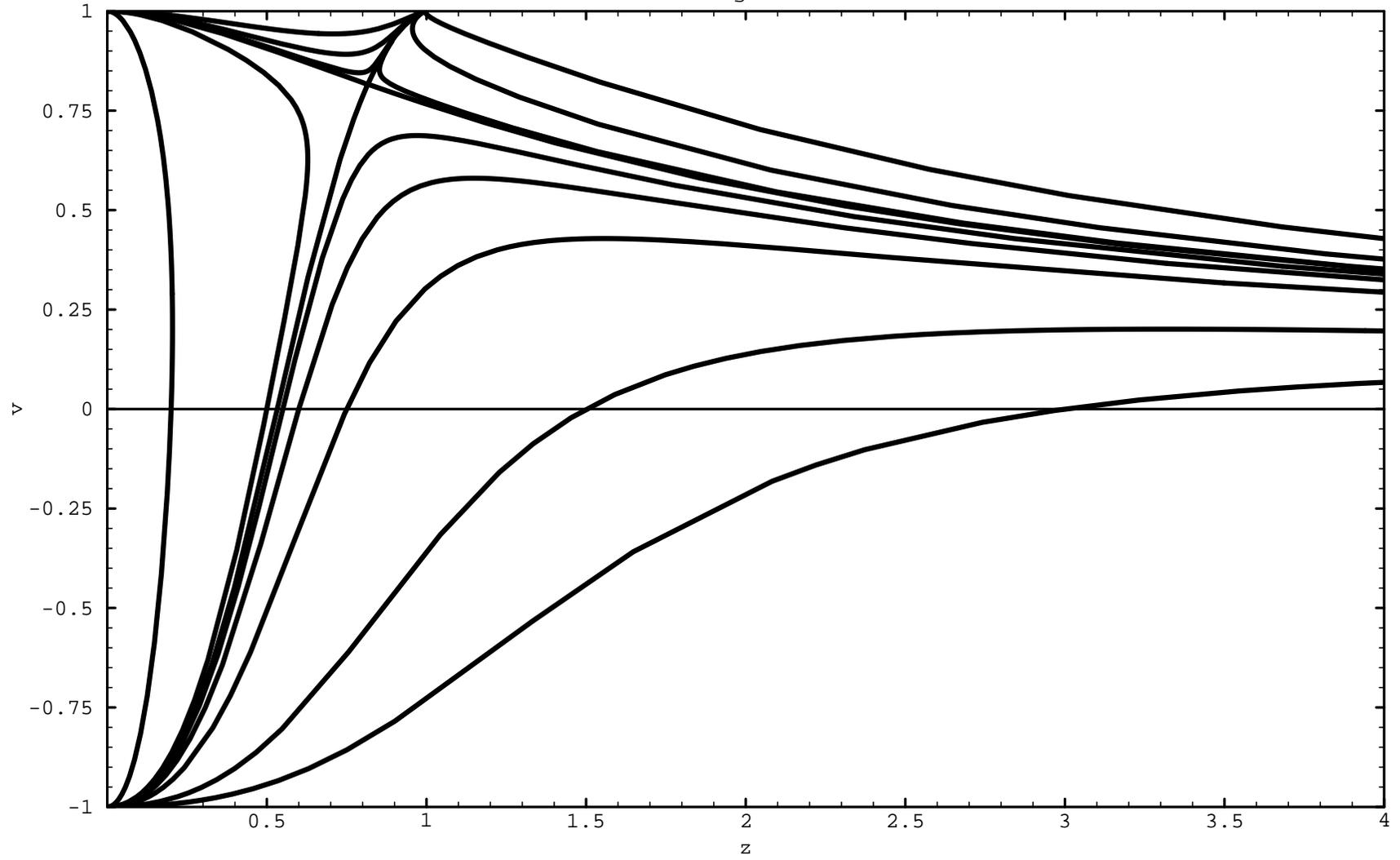
\begin{figure}
\caption{Physical strip of
the phase portrait of equation (\protect\ref{ecuared}) for $k=-3/2$.}
\label{fig2}
\end{figure}

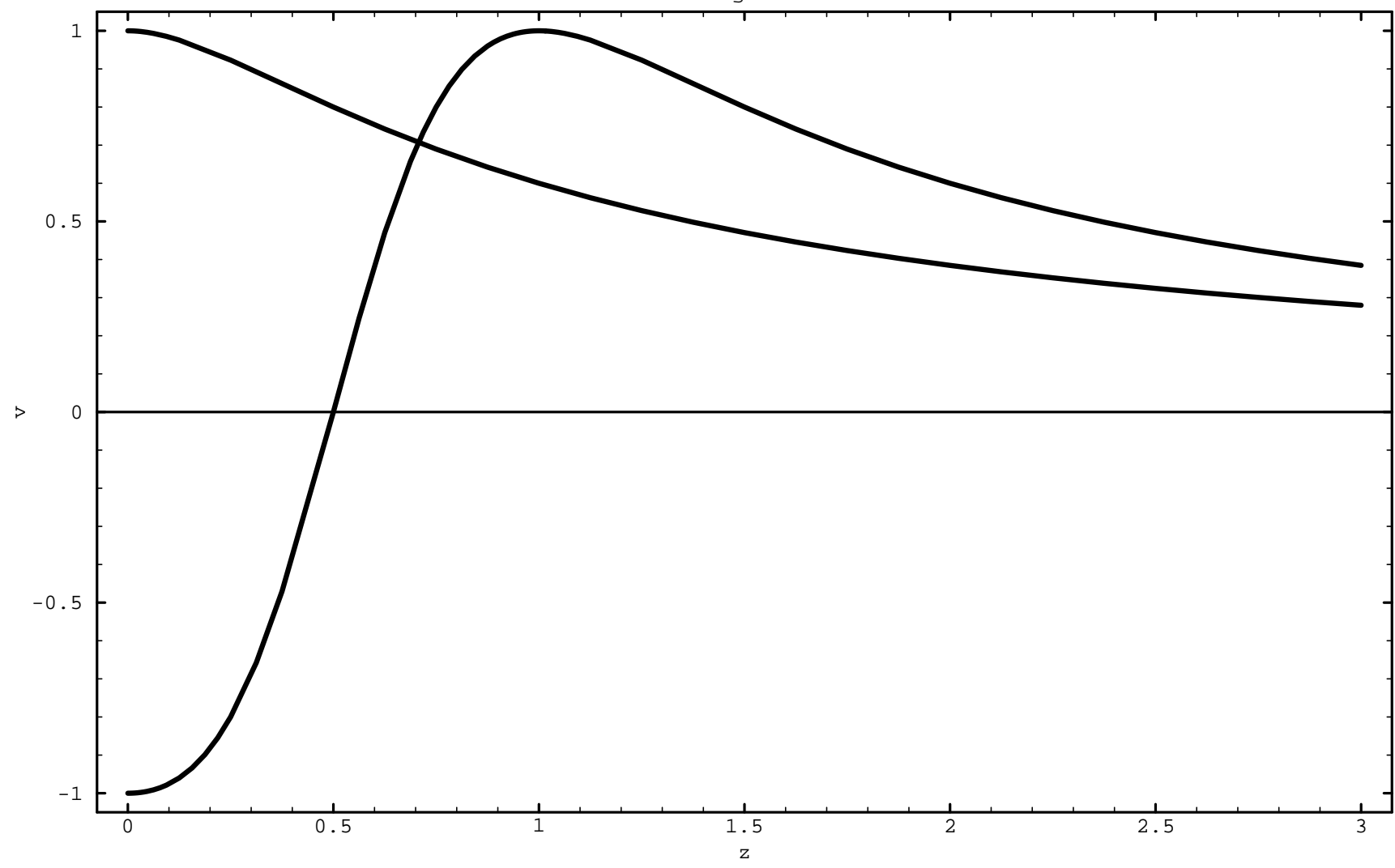
\begin{figure}
\caption{Separatrices for de Sitter spacetime}
\label{fig3}
\end{figure}

\begin{figure}
\caption{Two-string solution for $k=-0.5$: a) $|\eta|$ as a function of $\tau$;
b) $f$ as a function of $\eta$; c) $v$ as a function of $z$.}
\label{figdedos}
\end{figure}

\begin{figure}
\caption{Multistring solution for $k=-1.6$: a) $|\eta|$ as a function of
$\tau$;
b) $f$ as a function of $\eta$; c) $v$ as a function of $z$.}
\label{fig4}
\end{figure}
 
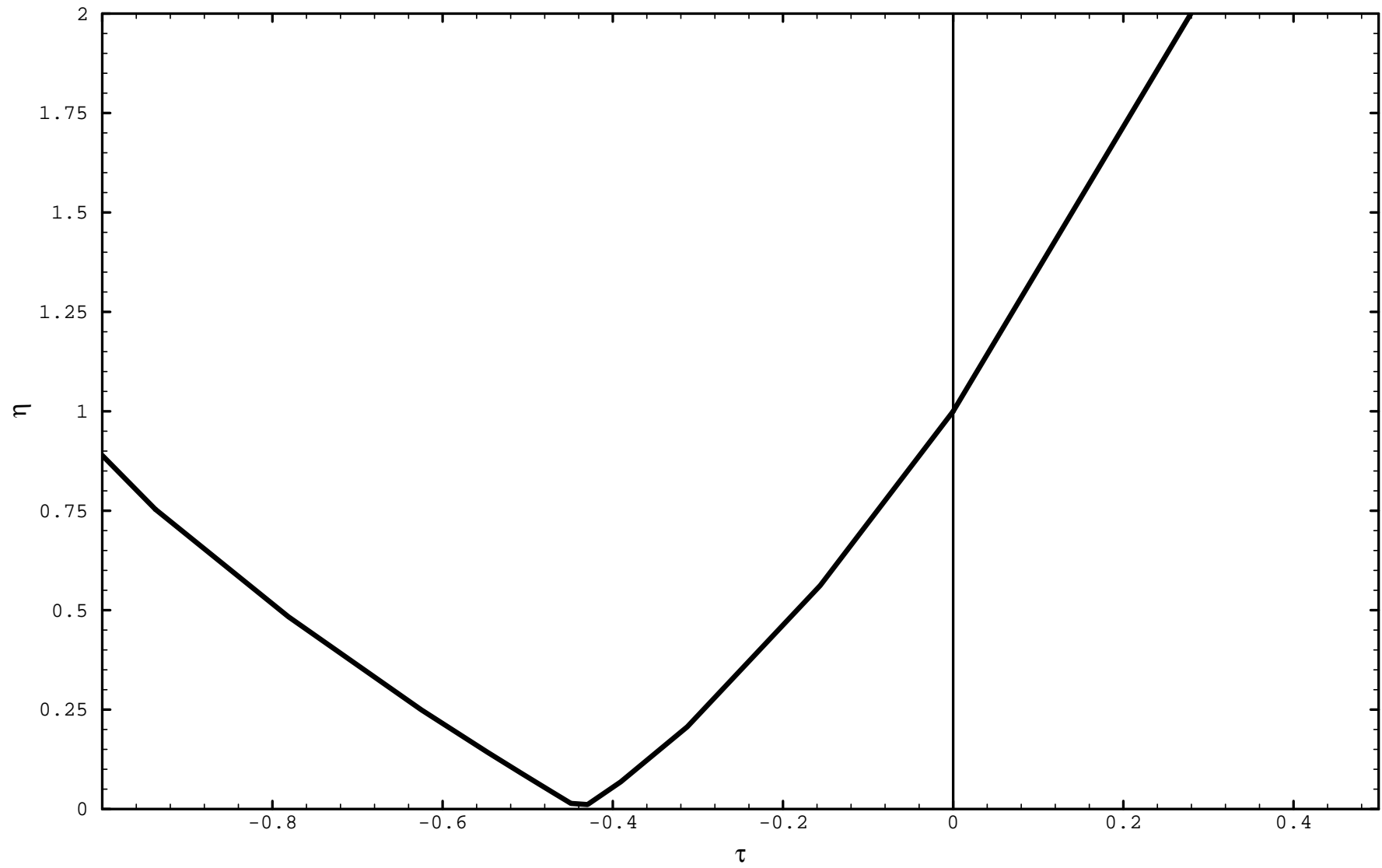
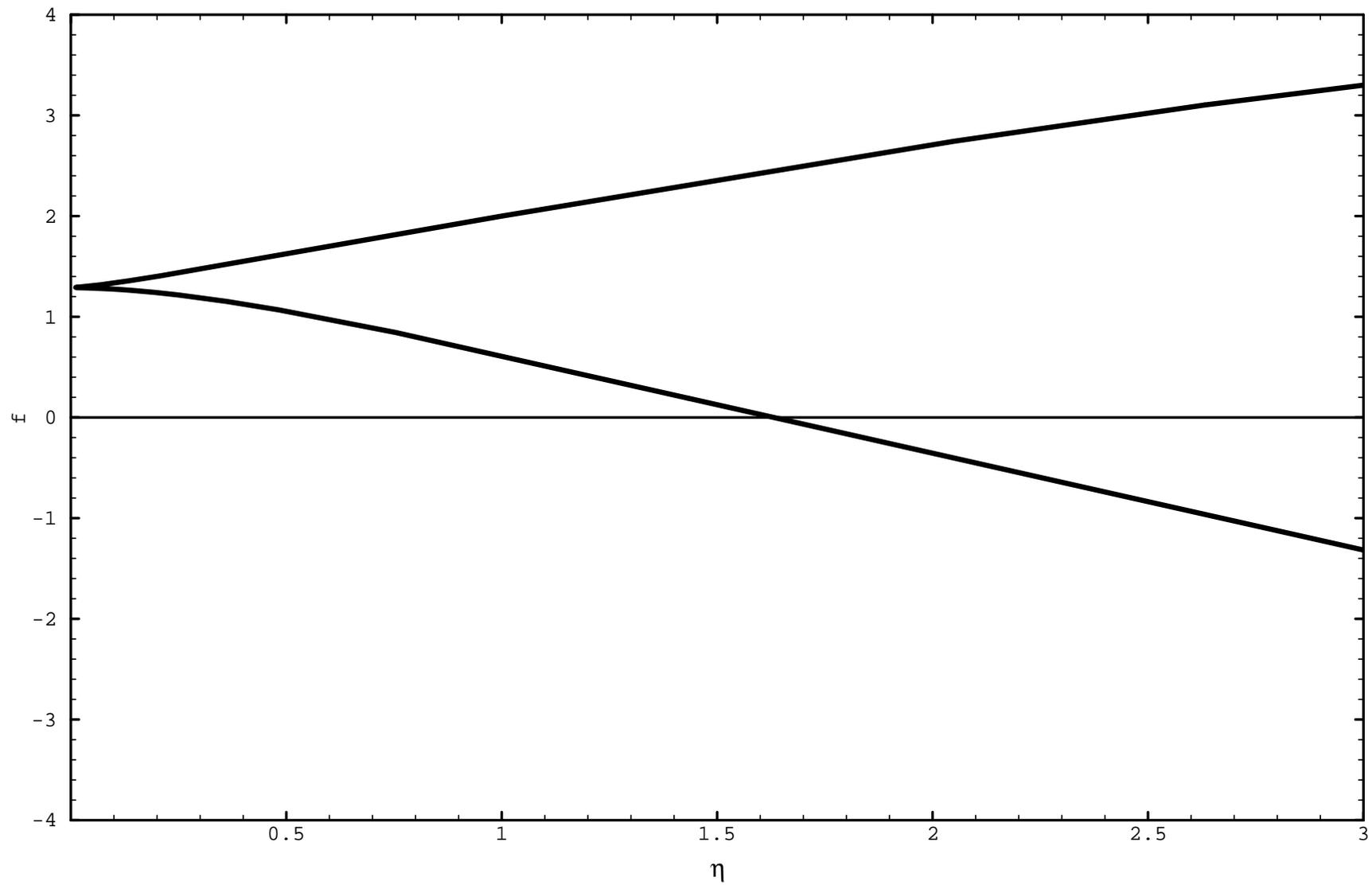
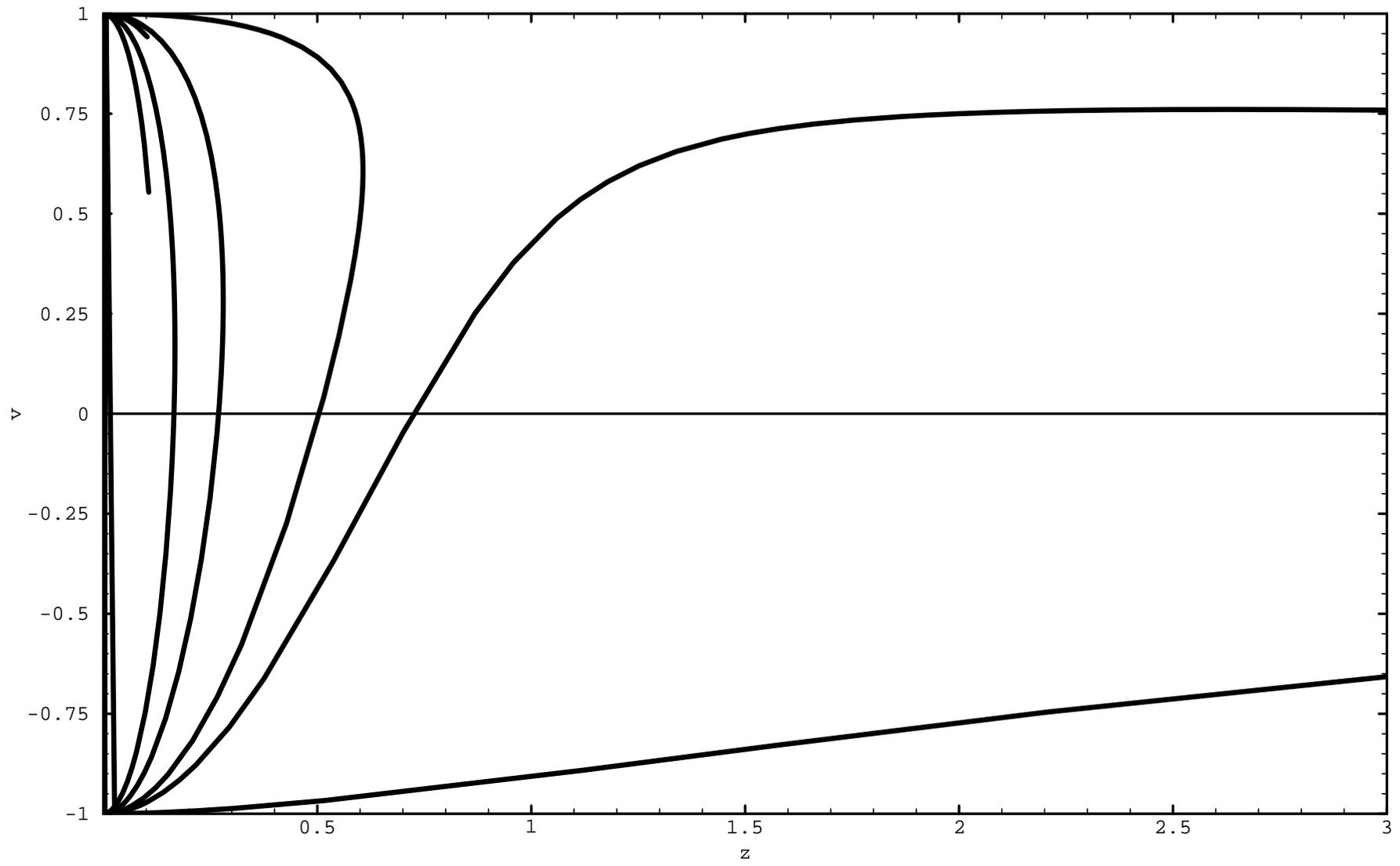
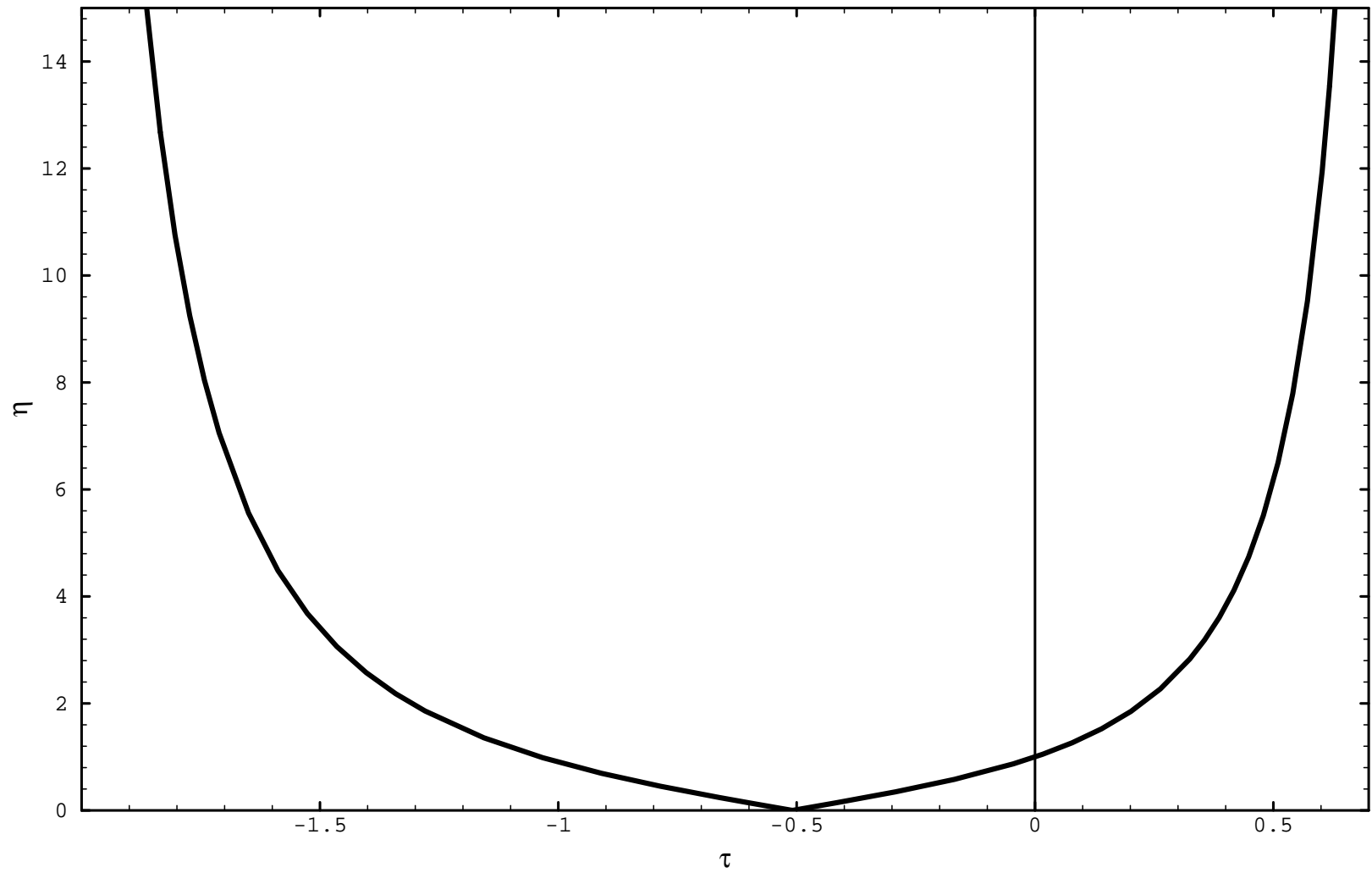
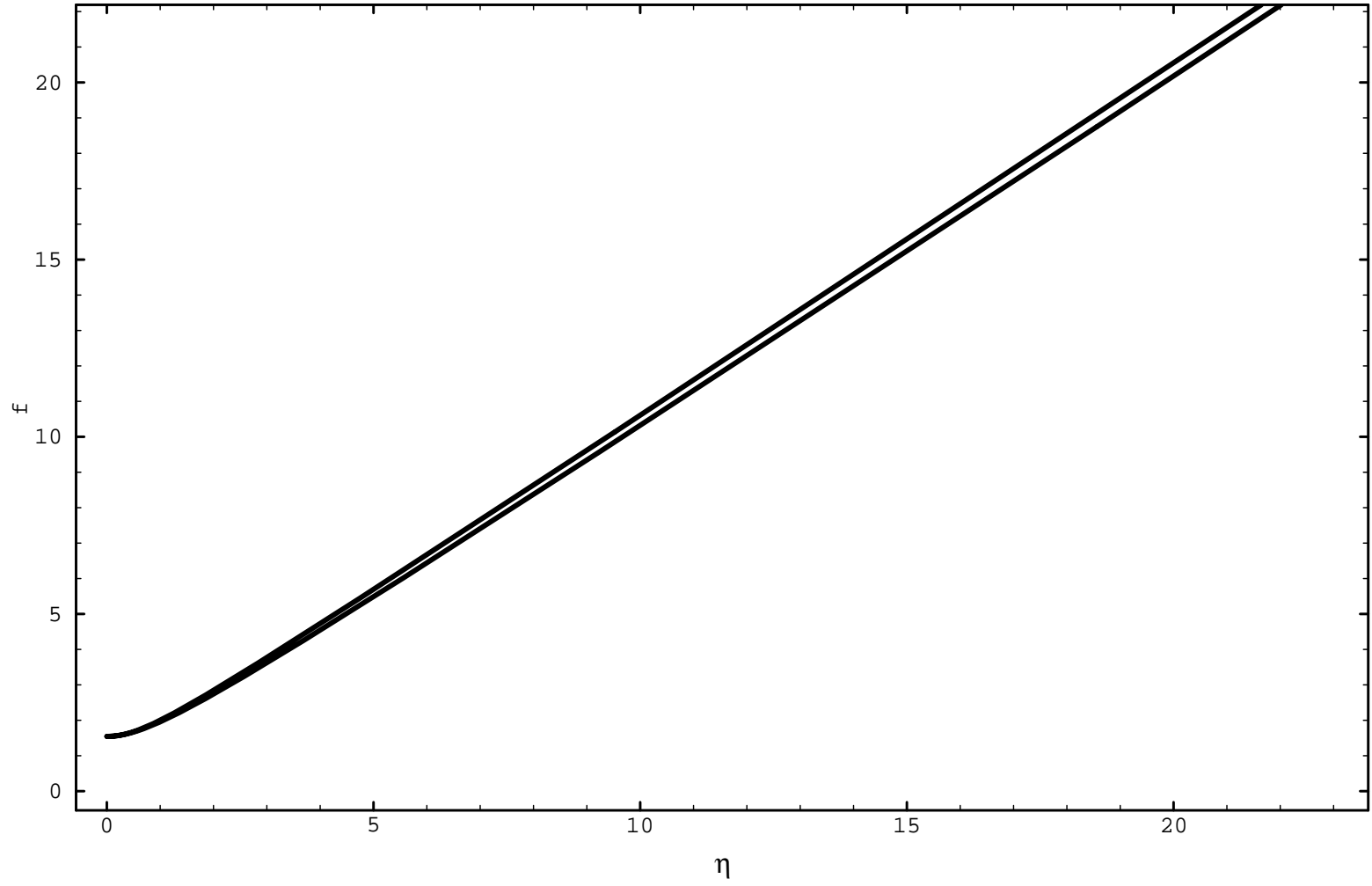
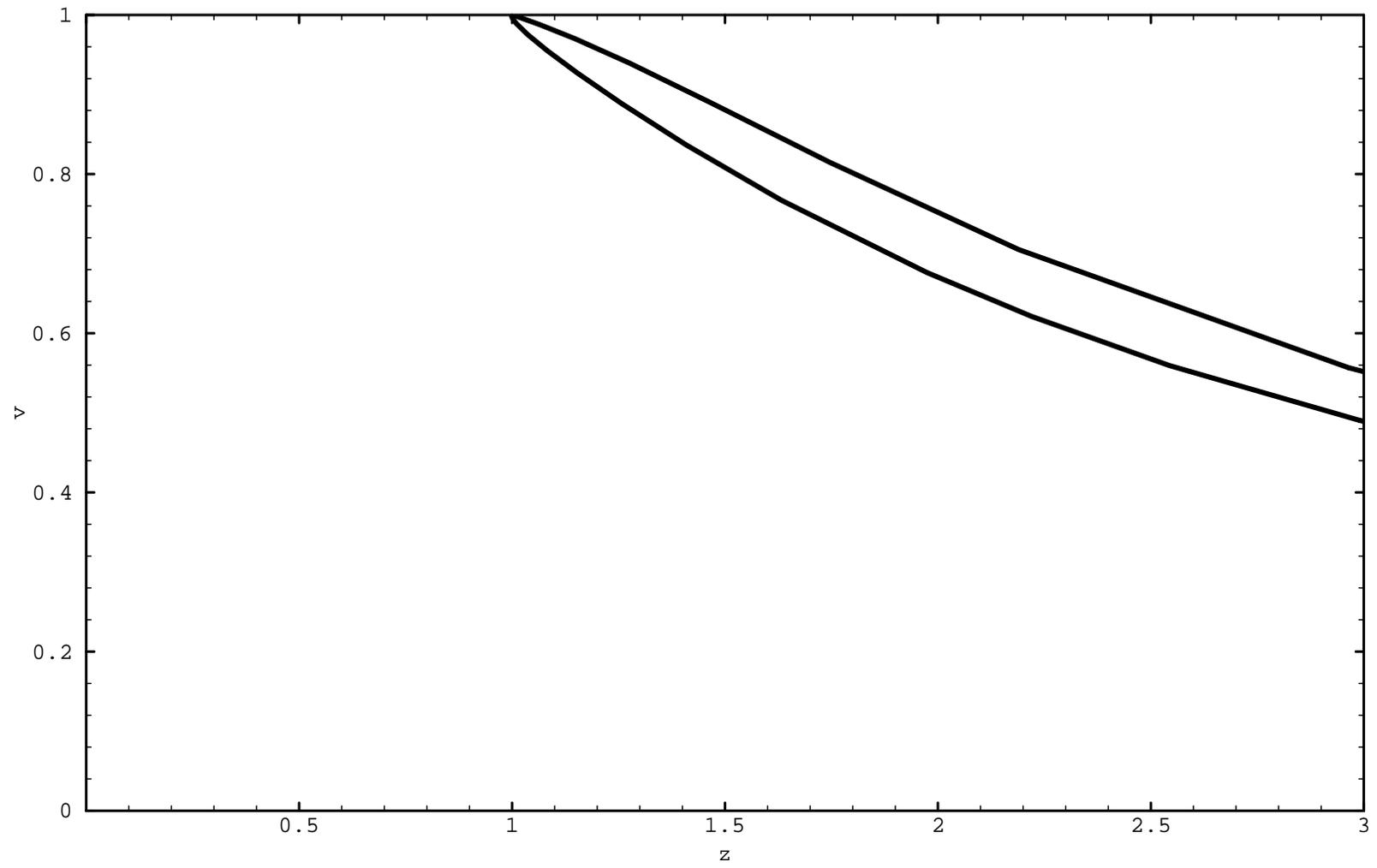
\end{document}